# Impact of Ce Substitution on Structural and Electrochemical Properties of Ga Doped Garnet $Li_7La_3Zr_2O_{12}$ Solid Electrolyte

Muktai Aote[1], A. V. Deshpande[1,*], Vaibhav Sirsulwar[1], Priya Padaganur[1], Neha[1], Abhishek Pradhan[1]

[1] *Department of Physics, Visvesvaraya National Institute of Technology, South Ambazari Road, Nagpur, 440010, Maharashtra, India*

[*]**Corresponding Author:**

**Dr. (Mrs.) A. V. Deshpande**

Department of Physics,

Visvesvaraya National Institute of Technology,

South Ambazari Road, Nagpur

Maharashtra, 440010 (India)

E-mail Address: avdeshpande@phy.vnit.ac.in

Ph.No.: 0-712-280-1173

## Abstract

In order to replace conventional liquid electrolytes, solid electrolyte should possess high ionic conductivity. In this study, the effects of Ga-Ce co-doping on the garnet $Li_7La_3Zr_2O_{12}$ solid electrolyte have been investigated. The series $Li_{6.4}Ga_{0.2}La_3Zr_{2-x}Ce_xO_{12}$ has been prepared with varying content of Ce from 0 to 0.30 atoms per formula unit (a.p.f.u.) by sintering at 1050ºC. Various structural characterizations namely X-diffraction, Scanning Electron Microscopy (SEM), density measurements were carried out. The electrochemical analysis suggested that, the sample with 0.10 a.p.f.u. of Ce offered the highest room temperature ionic conductivity of 4 x $10^{-4}$ S/cm with the minimum activation energy of 0.29 eV. Moreover, DC conductivity measurement proved the predominant ionic conduction in the prepared samples making it suitable for the application in all solid state Li-ion batteries (ASSLIBs).

**Keywords**

Li$_7$La$_3$Zr$_2$O$_{12}$ (LLZO); Ionic Conductivity; Garnet solid electrolyte; Activation Energy

# 1. Introduction

In the last few years, the market of portable electronic devices, electric vehicles and renewable energy storage devices has been developed extensively. This growth eventually relies on the safer, high energy and durable Li-ion batteries (LIBs), which has now put forward solid electrolyte as a most important component. Thus nowadays all the attention has been driven towards the investigation of Li-ion conducting solid electrolytes for all solid state Li-ion batteries (ASSLIBs) [1]. Despite having wide range of acceptance and great success in Li-ion batteries, conventional liquid electrolytes face extreme setbacks due to their various drawbacks including flammability, thermal and chemical instability and leakage prone behavior. Furthermore, these disadvantages of liquid electrolytes combined with their instability with Li metal anode. Thus, in order to enhance the safety as well as energy storage performance of Li-ion batteries, the adaption of solid electrolytes over organic liquid electrolytes is must [2].

The core element in the evolution of ASSLIBs from LIBs is the solid electrolyte, which must possess the high ionic conductivity at room temperature ($10^{-3}$-$10^{-4}$ S/cm) as that of liquid electrolytes along with the wide electrochemical potential window and chemical stability. Moreover unlike liquid electrolytes, solid electrolytes should have compatibility with lithium metal anode as well as high voltage cathode materials [3]. After the thought of replacing conventional electrolytes with solid electrolytes, different categories of solid electrolytes have been investigated and studied. These include sulfide solid electrolytes, polymer electrolytes and inorganic oxide electrolytes. However despite offering high conductivity, sulfide electrolytes could not get validation due to its moisture sensitive behavior along with environmental hazard. On the other hand, polymer electrolytes possesses excellent mechanical and chemical stability but can not be used in ASSLIBs due to their poor ionic conductivity at room temperature [4].

Similarly, wide range of inorganic oxide solid electrolytes, such as LIPON type, Perovskite type, LISICON type, and NASICON type have been studied. But each one of them faces the constraint in practical application either regarding the ionic conductivity or due to the lack of chemical and electrochemical stability [5]. However, Thangdurai et al. [6]discovered another category of solid electrolytes known as garnet type solid electrolytes which eventually surpasses all the other earlier mentioned electrolytes. In this garnet type, Murugan et al. [7] found the novel material identified as $Li_7La_3Zr_2O_{12}$ (LLZO), which later gained lot of attention due to its excellent properties. LLZO is stable against Li metal anode and offers wide electrochemical potential window (~6 V), which can be sustained against high voltage cathode materials. Moreover, LLZO offers great thermal and chemical stability with compatible ionic conductivity. In general LLZO has two crystal phases; namely tetragonal phase and cubic phase. Out of these two phases, tetragonal phase is thermodynamically stable at room temperature but offers less ionic conductivity (~$10^{-7}$ S/cm). In contrast, cubic phase has higher Li-ion conductivity (~ $10^{-3}/10^{-4}$ S/cm), but it is unstable at room temperature. Thus in order to use LLZO as a solid electrolyte in ASSLIBs, it must be stabilized in cubic phase [3,8].

To stabilize the cubic phase at room temperature various strategies have been adapted out of which interstitial doping is found to be very effective. In this case, the aliovalent/supervalent cations are doped at the site of Li, La and Zr in the lattice of LLZO [9]. Initially single doping technique has been utilized where various cations like, Al, Ge, Ta, Ca, Ba, Nb, Sr, Ga, etc. have been doped at any one of the site in LLZO [10–18]. After the successful experiments on single doped LLZO, recently multiple sites doping technique have been utilized where; two or more cations are doped simultaneously at the available sites in LLZO. It has been reported that, doping LLZO with supervalent cations resulted in the formation of Li-ion vacancies as well as stabilization of cubic phase [3,5,19]. This altogether led to the enhancement in Li-ion conductivity at room temperature. However, the transition of cubic phase from tetragonal phase needs higher sintering temperature, which eventually causes the lithium loss and results into lower ionic conductivity. Thus researchers are now focusing on the supervalent cations which can act as stabilizing agents as well as sintering aids.

From the literature, it is evident that doping of Ga at the Li site helps in the formation of Li-ion vacancies along with the stabilization of cubic phase. This will help in the increase in Li-ion conductivity at room temperature. Moreover, there is no much distortion of the lattice of garnet LLZO from the original crystal structure, as the ionic radius of Ga is comparable to that of Li [19,20]. On the other hand, literature supported the fact that, insertion of Ce at the Zr site in LLZO could be useful, as Ce will act as a sintering aid and help in the grain growth. This will lead to the densification of garnet LLZO which in turn enhances the Li ion migration pathways and reduces the chances of dendrite growth [21]. Although there are various studies reported on the investigation of single Ga doped LLZO, but there are very few studies reported on the effect of Ce on garnet LLZO. Some of the results regarding Ce doped LLZO have been tabulated in Table 1.

While individual doping of Ga and Ce have been explored earlier, the study on Ga-Ce co-doped LLZO is not done yet. Thus in the present study the synergistic effect of Ga-Ce dual doping on garnet LLZO has been investigated. Here, Ga and Ce have been doped at the site of Li and Zr respectively. However, it has been reported that, the excess of Ga content at the Li site decreases the overall lithium content and results into lower Li ion conductivity. Also, Ga tends to segregate into grain boundaries which also limit the ionic conductivity by offering grain boundary resistance [19]. Hence, in the current study Ga content is kept constant at 0.20 atoms per formula unit (a.p.f.u.), while the content of Ce has been varied from 0 to 0.30 a.p.f.u. in the garnet LLZO and the series $Li_{6.4}Ga_{0.2}La_3Zr_{2-x}Ce_xO_{12}$ was synthesized followed by its structural and electrochemical characterizations.

**Table 1: Various Compositions of Ce doped LLZO with the total ionic conductivity at room temperature.**

| Composition | Synthesis Method | Sintering Temperature &time | Total Ionic conductivity (S/cm) | References |
|---|---|---|---|---|
| $Li_7La_{2.5}Ce_{0.5}Zr_{1.625}Bi_{0.3}O_{12}$ | Sol Gel | 1150°C/ 6h | $5.12 \times 10^{-4}$ | [21] |
| $Li_{6.6}La_{2.6}Ce_{0.4}Zr_2O_{12}$ | - | - | $1.44 \times 10^{-5}$ | [22] |
| $Li_7La_3Zr_{1.75}Ce_{0.25}O_{12}$ | Solid State Reaction | 1250- 1150°C | $2.2 \times 10^{-4}$ | [23] |
| $Li_{6.5}La_3Zr_{1.375}Nb_{0.5}Ce_{0.125}O_{12}$ | Solid State Reaction | 1200°C/ 10h | $\sim 7 \times 10^{-4}$ | [24] |
| Li$_7$La$_3$Zr$_{1.75}$Ce$_{0.25}$O$_{12}$ | - | 1000- 1050ºC/ 12h | 1.2 x 10$^{-4}$ (50ºC) | [25] |

## 2. Experimental Work

### 2.1. Material Synthesis

The series Li$_{6.4}$Ga$_{0.2}$La$_3$Zr$_{2-x}$Ce$_x$O$_{12}$ was prepared by conventional solid-state reaction method with Ce (x) ranging from 0 to 0.30 a.p.f.u. The chemicals, namely, Li$_2$CO$_3$ (Merck, >99.9%), Ga$_2$O$_3$ (Sigma Aldrich, >99.0%), La$_2$O$_3$, ZrO$_2$, and CeO$_2$ (Sigma Aldrich, >99.99%) were weighed stoichiometrically and mixed in agate mortar. During the initial mixing process, 10% of excess Li$_2$CO$_3$ was added to compensate for the Li loss, which occurs during the sintering process. After mixing, the powder was calcined at 900ºC for 8 h in a muffle furnace. Once the calcination was done, the powder was again grounded into fine particles. The pellets with diameter of 10 mm and thickness of around 1.5 mm were made using a hydraulic press under pressure of 4 tons. The pellets were then kept in the mother powder bed and sintered at 1050º C for 8 h. The prepared samples of the series Li$_{6.4}$Ga$_{0.2}$La$_3$Zr$_{2-x}$Ce$_x$O$_{12}$ are represented as 0 Ce, 0.10 Ce, 0.20 Ce and 0.30 Ce , for Ce(x) content varying from 0, 0.10, 0.20 and 0.30 respectively.

### 2.2. Characterization Techniques

Various characterizations have been carried out to investigate the effect of Ga and Ce co-doping on the structural and electrical behavior of garnet LLZO. For the cubic phase identification, the synthesized pellets were grounded into a fine powder and then examined through a RIGAKU X-ray diffractometer. The sample was exposed to Cu-kα radiation which has a wavelength of 1.52 Å. The required data was collected in the range of 10º - 70º by keeping a scan speed of 2º/ min with a step size of 0.02º. Archimedes' principle was used to calculate the densities of synthesized samples using the K-15 Classic K-Roy balance. During density measurement, toluene was used as an immersion medium. The information regarding the surface morphology and elemental composition were investigated through Scanning Electron Microscopy (SEM) using the JSM-7600F/JEOL instrument. The electrochemical analysis of the samples was done using NOVOCONTROL impedance analyzer, where AC conductivity measurements were carried within the frequency range of 20 Hz to 20 MHz with temperature ranging from room temperature to 150ºC. Furthermore, the DC polarization technique was employed using KEITHLEY 6512 programmable electrometer to ensure the predominance of ionic conduction in the prepared samples. The ionic transport number was calculated using the data of DC conductivity. For the electrochemical measurements, the surface of sintered pellets was coated with silver paste to maintain ohmic contact with the silver electrodes. These electrodes functioned as ion-blocking electrodes in these measurements.

## 3. Results and Discussion

### 3.1. X-Ray Diffraction Analysis

The X-ray diffraction patterns of all the synthesized samples of the series $Li_{6.4}Ga_{0.2}La_3Zr_{2-x}Ce_xO_{12}$ (x=0 to 0.30) are given in Fig.1 (a). The formation of conducting cubic phase (Ia-3d) in garnet LLZO has been confirmed from the JCPDS file no 45.0109. From Fig.1 (a), it can be observed that, all the samples possess the cubic phase and the respective miller indices (h k l) have been indicated. The vertical black lines below the plots correspond to the data of pure LLZO. Previously, Rangasamy et al.[22] reported that for the stabilization of cubic phase in garnet LLZO, Ce content must be greater than 0.20 a.p.f.u.. However, in present study the required cubic phase has been achieved at much lower Ce content. This can be attributed to the insertion of Ga along with Ce in the lattice of LLZO, which helped in the stabilization of cubic phase. Moreover, it can be observed that, with the increase in concentration of Ce, the peak intensity is also increased. This can be due to the crystallization in LLZO owing to the sintering ability of Ce. Hence, it can be stated that, Ga-Ce co-doping in garnet LLZO helped in the formation of cubic phase without any probable peak splitting which generally occurs due to the presence of secondary phase. However, $La_2Zr_2O_7$ namely pyrochlore phase can be seen which often resulted due to the Li evaporation during sintering [5]. The successful insertion of Ce in Ga doped LLZO can be confirmed from the shifting of peaks as depicted in Fig. 1(b). With the increase in content of Ce, the peaks are shifted towards the lower theta value. This can be due to the substitution of higher ionic radii Ce (0.87 Å) at the site of lower ionic radii Zr (0.78 Å). This shifting of peak towards lower theta value also suggested the possible expansion of lattice. The lattice constant for all the synthesized samples has been derived from the Reitveld refinement and the obtained results are shown in Fig.2. Initially for 0 Ce, the lattice constant has been reduced from 13.0035 Å (pure LLZO) to 12.9814 Å [23]. However, with the increase in Ce content, the lattice constants are increased. As the Ce content increases beyond the optimum limit of 0.20 a.p.f.u., there is decrease in lattice constant. The same result can be correlated with the right shifting of peak for 0.30 Ce. This can be due to the fact that, after optimum limit of Ce doping, it can not be substituted in LLZO lattice and may reside in the region of grain boundaries.

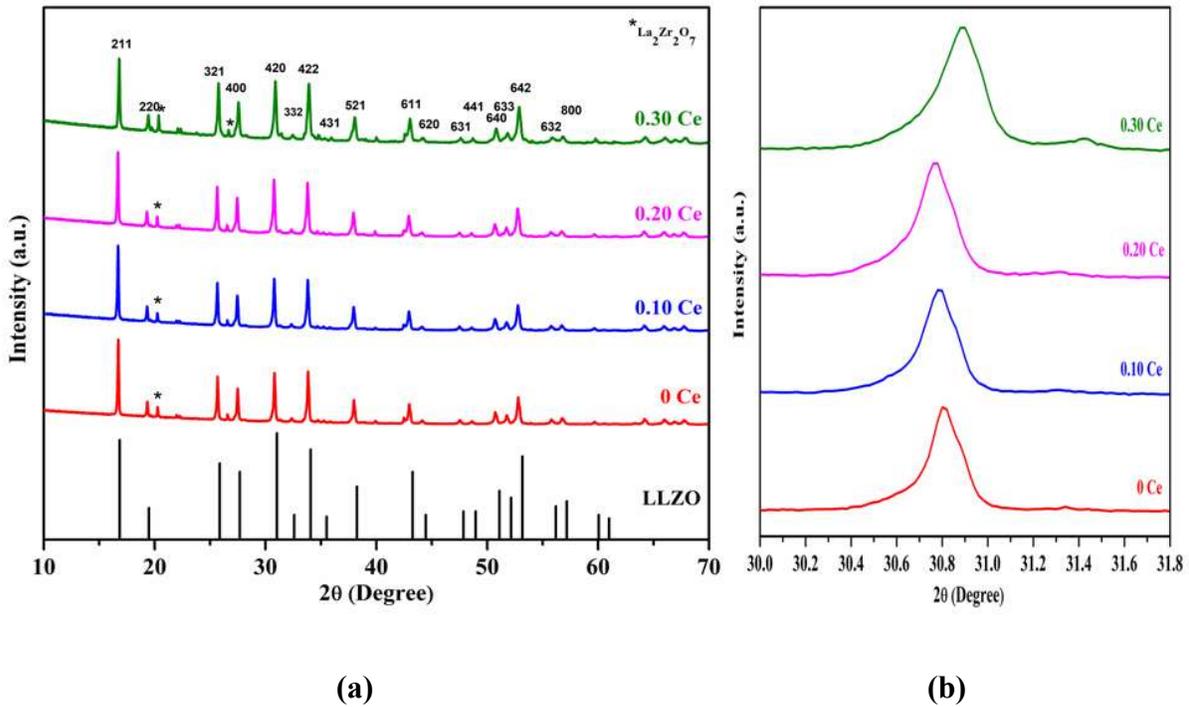

(a)            (b)

**Fig.1: (a) X-ray diffraction patterns of all the samples of series $Li_{6.4}Ga_{0.2}La_3Zr_{2-x}Ce_xO_{12}$ (x=0 to 0.30) (b) Shifting of (3 2 1) peak.**

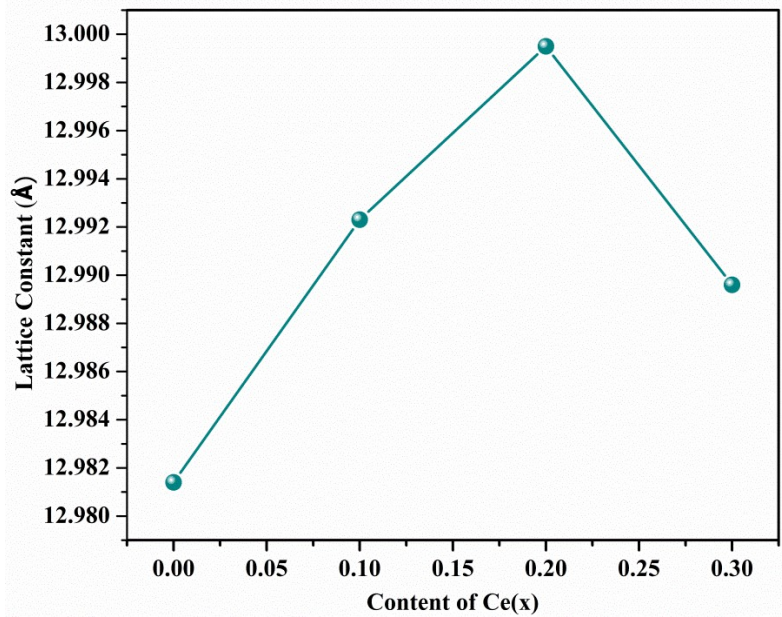

**Fig.2: Variation in lattice constant with the content of Ce(x) in $Li_{6.4}Ga_{0.2}La_3Zr_{2-x}Ce_xO_{12}$.**

### *3.2. Density Measurement Analysis*

For application in ASSBs, the mechanical strength of solid electrolyte is often estimated from various parameters including relative density [26,27]. In the current study, the densities of the synthesized samples of $Li_{6.4}Ga_{0.2}La_3Zr_{2-x}Ce_xO_{12}$ (x=0-0.30) series were calculated using Archimedes' principle and toluene was used as an immersion medium. The respective relative densities of all the samples are plotted against the content of varying Ce and shown in Fig.3. From the figure it can be observed that, with the initial addition of 0.10 Ce the relative density was increased as compared to 0 Ce, and found to be maximum of 95.96% with experimental density of 4.79 g/cm$^3$. The obtained density value is greater than single Ce doped LLZO, where relative density of 94% was obtained despite of using hot press sintering. Moreover, the obtained highest relative density (> 95%) is in well accordance with the previous studies where Ce is doped with Bi and Ta in garnet LLZO [21,28]. Owing to the synergistic effect of dual doping strategy, this can be attributed to the substitution of Ce as a sintering additive in garnet LLZO which leads to the shrinkage of garnet lattice and resulted into compact and dense structure [21]. However, with the further increase in Ce content beyond 0.10 a.p.f.u., the decrease in relative density can be observed. As reported in earlier studies, Ce reached the solubility limit beyond (>0.20 a.p.f.u.) [22]. This might be the reason of declined densities after 0.10 Ce. Also, the surface morphology also affects the density which can be seen in the following section.

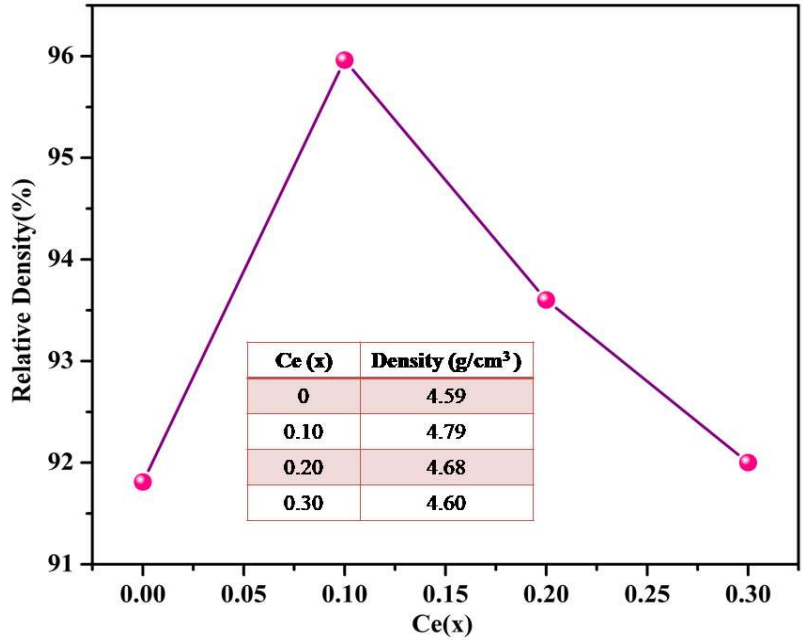

**Fig.3: Variation of density and relative density of $Li_{6.4}Ga_{0.2}La_3Zr_{2-x}Ce_xO_{12}$ with Ce(x).**

*3.3. Surface Morphological Analysis*

The surface morphology study has been carried out for all the samples. Fig. 4 (a-d) shows the scanning electron microscopy (SEM) images of the samples of $Li_{6.4}Ga_{0.2}La_3Zr_{2-x}Ce_xO_{12}$ series with Ce(x) content ranging from 0 to 0.30 a.p.f.u.. Fig. 4(a) corresponds to the surface micrograph of 0 Ce sample. From the figure it can be observed that, there is irregular growth of grains within the sample with voids between the grains. This eventually leads to the minimum density value as mentioned in earlier section. However, as the Ce gets inserted into the lattice there is uniform grain growth. Fig. 4(b) for 0.10 Ce sample shows the compact and dense microstructure. Also, the grains are very well connected with the neighboring grains giving the single crystal like structure. Also, no pores and voids can be seen in the surface micrograph of 0.10 Ce sample. This result is in contradiction with the previously reported studies, where even after doping the Ce at highest amount of 0.10 a.p.f.u., the pores were not reduced [28]. Hence, the obtained result in present study can be attributed to the effect of Ga-Ce co-doping. Because, even though Ga helped in the formation of larger grains [29], the Ce incorporation led to the merging of grains which suggests very well sintering behavior of Ce. This is also supported by the maximum relative density for 0.10 Ce sample as mentioned in table 2. The average particle size was calculated for 0.10 Ce sample and shown in the Fig. 5 with the corresponding particle size distribution histogram. The average particle size was found to be 7 µm for 0.10 Ce which is in well accordance with the result reported by Wang et al. [21]. The highest relative density with large grain size gives the advantage to 0.10 Ce sample in the ionic conduction of Li-ions [3]. However, with the further increase in Ce content beyond 0.10 a.p.f.u., there is formation of irregular cluster structure having grains sticking together and there is non-uniformity in the grain size. This can be observed from Fig. 4(c-d). Also in the surface image of 0.30 Ce, the inter particle growth of small grains within the large grain can be observed in the form of irregular polyhedral blocks. Furthermore, this irregular grain growth resulted into the creation of large voids within the structure. As the void between these grains is large, it has directly affected the relative density of the sample and thus the ionic conductivity. The presence of all the constituent elements in the series $Li_{6.4}Ga_{0.2}La_3Zr_{1.9}Ce_{0.1}O_{12}$, namely La, Zr, Ga, Ce and O have been confirmed from the elemental mapping as shown in Fig.6. From the Fig., it can be clearly seen that, all the mentioned elements are uniformly distributed over the surface of 0.10 Ce sample except Li.

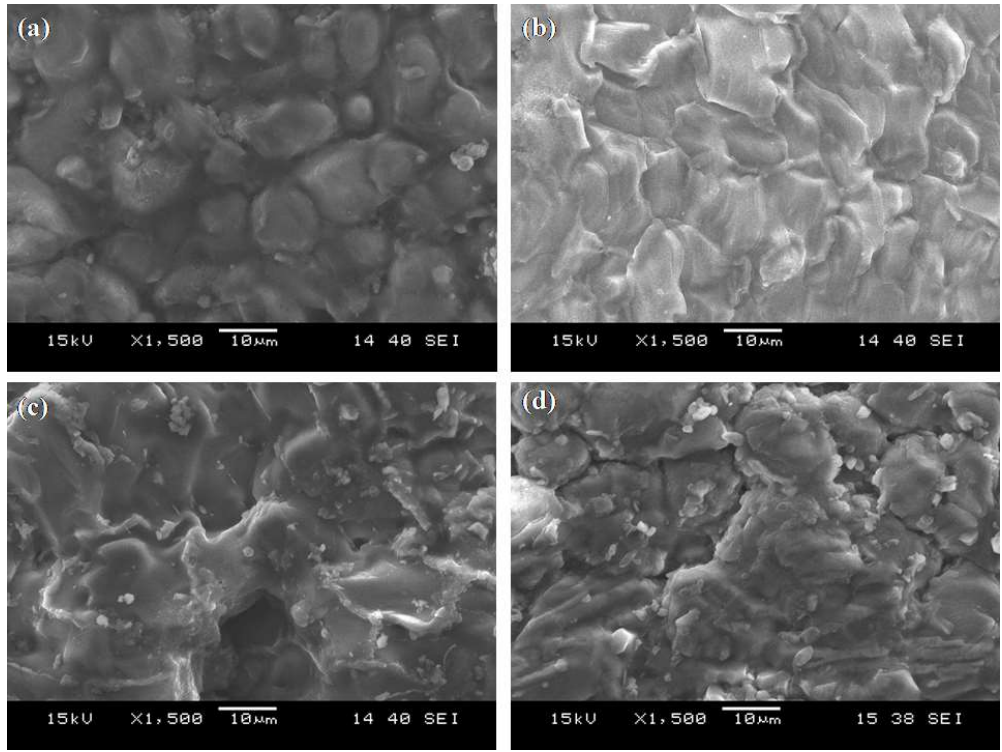

**Fig.4: SEM images of $Li_{6.4}Ga_{0.2}La_3Zr_{2-x}Ce_xO_{12}$ with x = a) 0, b) 0.10, c) 0.20, d) 0.30.**

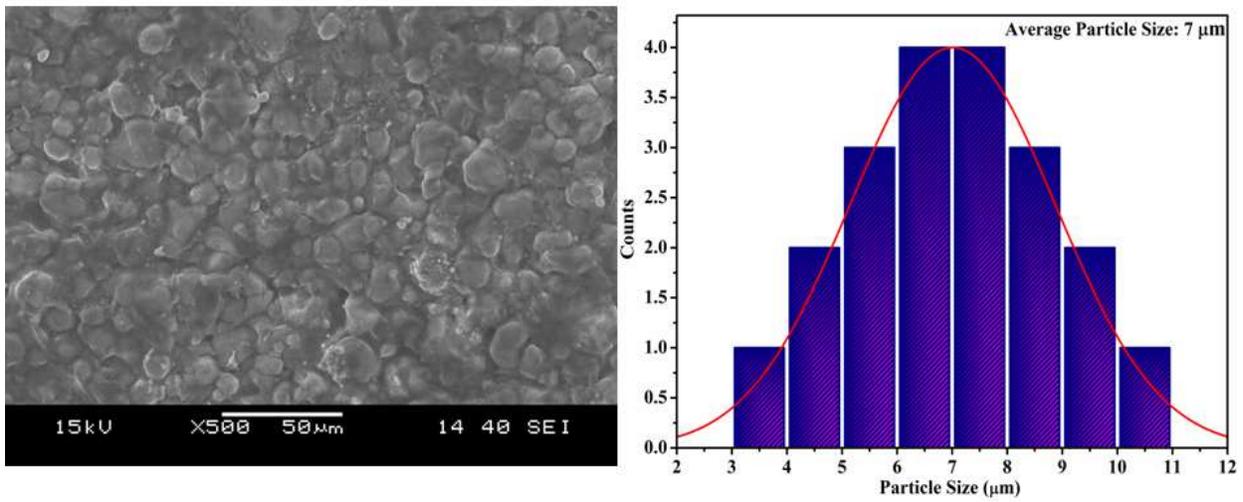

**Fig. 5: Average particle size distribution histogram of 0.10 Ce sample.**

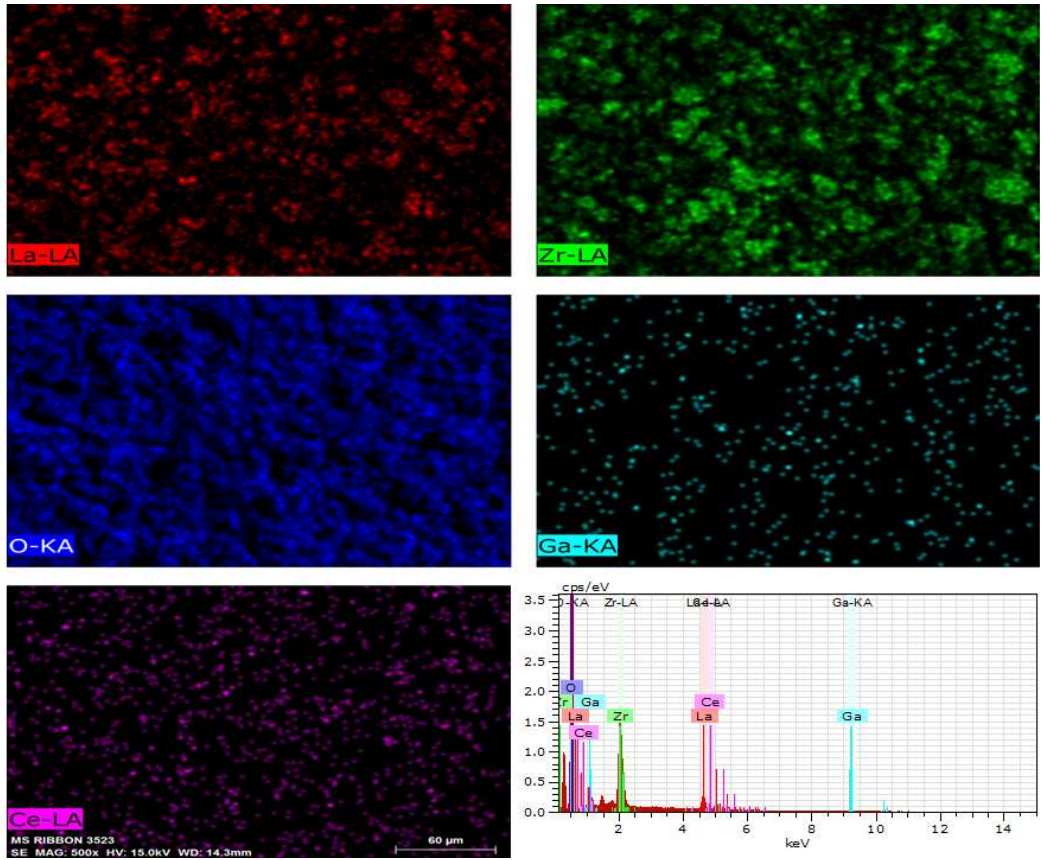

**Fig.6: Elemental mapping with EDS spectra of 0.10 Ce sample.**

*3.4. AC Conductivity Analysis:*

*3.4.1. Ionic Conductivity Analysis*

To get an idea about the ionic conductivity offered by each sample in the series $Li_{6.4}Ga_{0.2}La_3Zr_{2-x}Ce_xO_{12}$ (x= 0-0.30), the complex impedance study has been carried out and the respective nyquist plots are plotted at room temperature and shown in Fig.7 (a). The basic nature of nyquist plot contains the semicircle at high frequency along with the tail in the lower frequency region. The presence of tail corresponds to the Li-ion blocking nature of the electrodes. Here, impedance of the sample can be determined by the intercept made by the semicircle on the real Z axis. From the Fig.7 (a), it can be seen that for each sample of the series $Li_{6.4}Ga_{0.2}La_3Zr_{2-x}Ce_xO_{12}$, only one semicircle can be observed. This is because the fact that, at higher frequency range, it is almost difficult to separate the resistance offered by grain and grain boundary separately. Thus, the bulk resistance can be calculated from the obtained impedance spectra. The room temperature ionic conductivity was calculated using the formula $\sigma_{total} = t/RA$, where, $\sigma_{total}$, $t$, $R$ and $A$ represent the ionic conductivity, thickness, resistance and the area of the individual samples respectively. From Fig.7 (a), it can be observed that, the 0 Ce sample has the maximum intercept on the real Z axis offering the maximum impedance among all prepared samples. Whereas, the impedance decreases with the insertion of Ce. Moreover, out of all the samples, the 0.10 Ce sample shows the minimum intercept owing to maximum ionic conductivity of 4 x $10^{-4}$ S/cm. The fitted nyquist plot with equivalent circuit for 0.10 Ce sample is shown in Fig. 7(b), where $R_1$ and $R_2$ correspond to the resistances of grain and grain boundary respectively. This obtained ionic conductivity for 0.10 Ce is greater than earlier reported study on single Ce doped LLZO, in which the highest content of Ce (~0.4 a.p.f.u) could not achieve the desirable conductivity [22]. Thus, the obtained maximum value of ionic conductivity for 0.10 Ce sample in the present study can be attributed to the combined effect of Ga-Ce co-doping in the garnet LLZO. Here, Ga helped in the stabilization of conducting cubic phase and the insertion of Ce helped in achieving the highest relative density, which ultimately increases the Li-ion migration pathways. Also, it is reported that, the Ce incorporation in LLZO resulted into the re-distribution of Li-ions which in turn affects the Li-ions conductivity [30]. However, as the Ce content increases further 0.10 a.p.f.u., the semicircles in the nyquist plots shifted towards high impedance region, resulted into the lowering of ionic conductivity. This can be due to the formation of voids and inter-granular structure which obstructs the Li ion migration within the structure. Moreover, the decrease in ionic conductivity values is in well agreement with the study reported by Ziqiang Xu et al.[28], where, the lowering of ionic conductivity was observed for > 0.10 Ce a.p.f.u. content. Thus from the current study it can be confirmed that, 0.10 a.p.f.u. is the optimum content of Ce in order to achieve the high room temperature ionic conductivity.

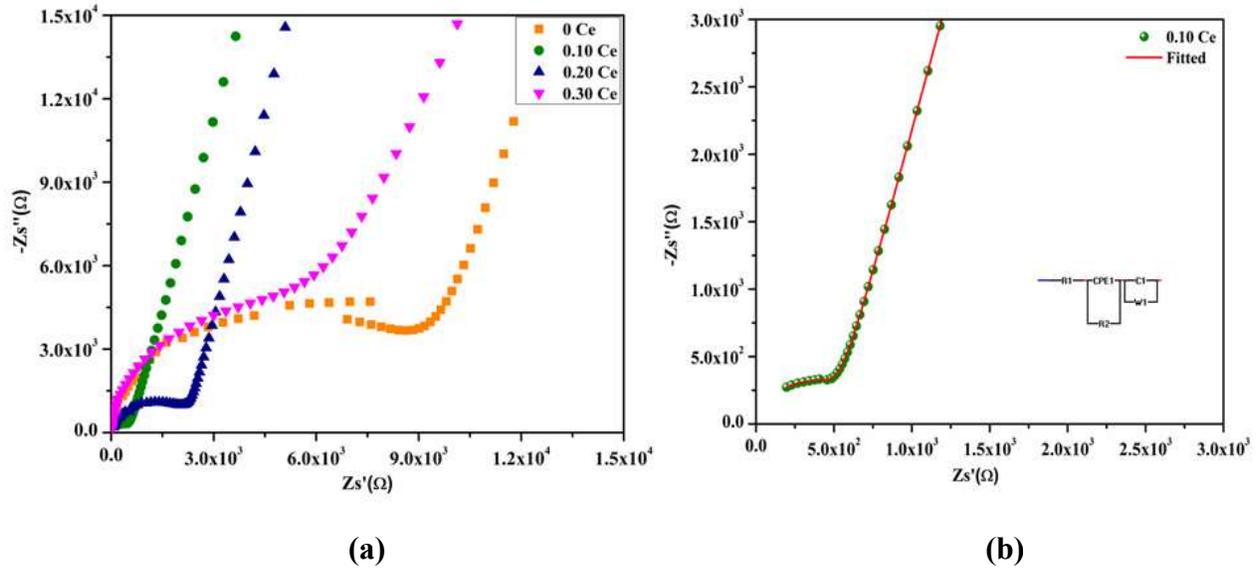

(a)  (b)

**Fig.7: a) Nyquist plots at room temperature for the series $Li_{6.4}Ga_{0.2}La_3Zr_{2-x}Ce_xO_{12}$ with varying content of Ce(x) from 0 to 0.30 a.p.f.u., b) Fitted spectra with equivalent circuit for 0.10 Ce sample.**

### 3.4.2. Arrhenius Plots Analysis

The temperature dependent nature of ionic conductivity can be studied from the Arrhenius equation. The Arrhenius plots for the samples of series $Li_{6.4}Ga_{0.2}La_3Zr_{2-x}Ce_xO_{12}$ are depicted in Fig. 8 (a) within the temperature range of 25ºC to 150ºC. From the graph it can be clearly seen that, the ionic conductivity values have been increased with the increase in temperature, and thus follow the Arrhenius equation. Also, this linear equation obtained for all the synthesized samples suggest that, there is no structural or phase change that occurred during the high temperature treatment. Thus it can be confirmed that, the synthesized series is thermally stable. The activation energy is calculated using the Arrhenius equation as $\sigma(T) = \sigma_0 exp(-E_a/K_BT)$, where $\sigma$ is the conductivity, $\sigma_0$ is the pre-exponential factor, $E_a$ is the activation energy, $K_B$ is the Boltzmann constant, and $T$ is the temperature in Kelvin. The ionic conductivity and activation energy have the inverse relation which can be observed from Fig. 8(a). Here, the minimum activation energy of 0.29 eV is obtained for 0.10 Ce sample which possesses the highest ionic conductivity of 4 x $10^{-4}$ S/cm. The values of ionic conductivity and respective activation energy are mentioned in table 3 for all the samples. Moreover, the variation of ionic conductivity at room temperature and activation energy with the varying content of Ce is shown in Fig. 8(b). This obtained result of activation energy for 0.10 Ce sample is slightly higher than the earlier Ce doped LLZO studies. This can be attributed to the insertion of supervalent Ce at the site of Zr which enhances the lattice and helpes in the tuning of bottleneck structure for Li ion migration by increasing the Li ion migration pathways [23,24]. Thus, the optimum doping of Ce along with Ga, which initially helped in the stabilization of cubic phase by creating the Li ion vacancies can be the suitable choice for enhancing the electrochemical performance in garnet LLZO.

**Table 2: The values of Ionic conductivity at room temperature and the activation energy of $Li_{6.4}Ga_{0.2}La_3Zr_{2-x}Ce_xO_{12}$**

| Content of Ce | Ionic Conductivity (S/cm) | Activation Energy (eV) |
|---|---|---|
| 0 | 2.41 x $10^{-5}$ | 0.44 |
| **0.10** | **4 x $10^{-4}$** | **0.29** |
| 0.20 | 9.93 x $10^{-5}$ | 0.34 |
| 0.30 | 4.26 x $10^{-5}$ | 0.41 |

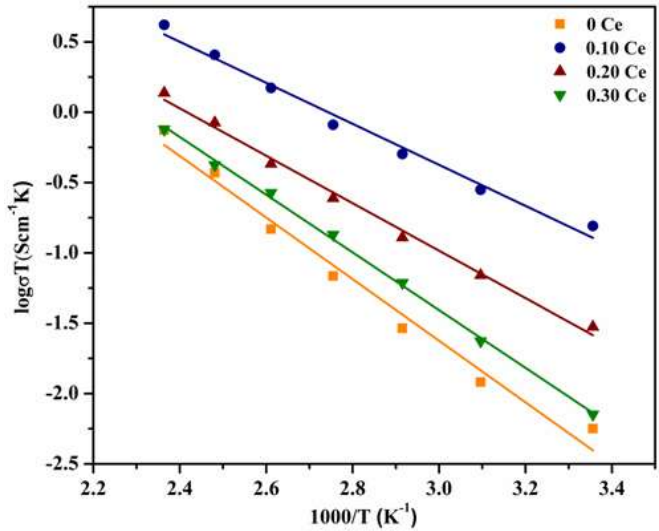

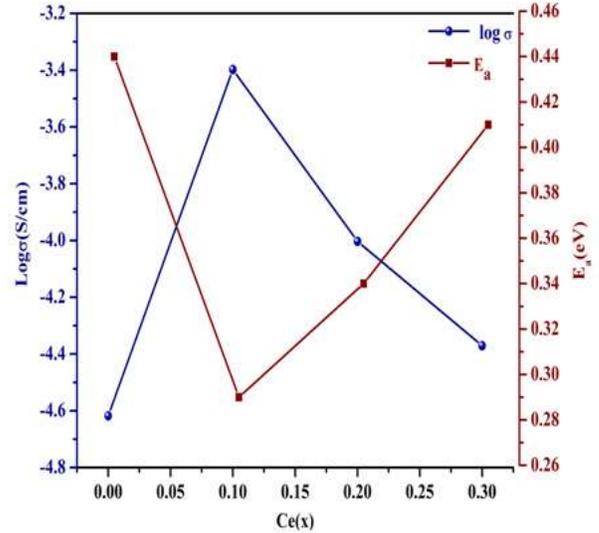

(a)　　　　　　　　　　　　　　　　(b)

**Fig.8: a) The Arrhenius plots and b) Variation in activation energy and ionic conductivity for $Li_{6.4}Ga_{0.2}La_3Zr_{2-x}Ce_xO_{12}$ series.**

### 3.5. DC Conductivity Analysis

For the confirmation of nature of conduction in synthesized samples of the series $Li_{6.4}Ga_{0.2}La_3Zr_{2-x}Ce_xO_{12}$, the DC conductivity measurements have been carried out and the DC conductivity plot for the highest conducting 0.10 Ce sample is shown in Fig.9. During the measurement the constant voltage of 1 V was applied and the current was measured as a function of time. The ionic transport number was calculated using the formula, $t_i = (\sigma_{total} - \sigma_e)/\sigma_{total}$. From the Fig., it can be observed that, after short interval of time the current through the sample remained constant and this current is only due to the electrons. Moreover, the ionic transport number calculated for 0.10 Ce sample was found to be > 0.999, which shows the predominance of ionic conduction within the sample [3–5][31].

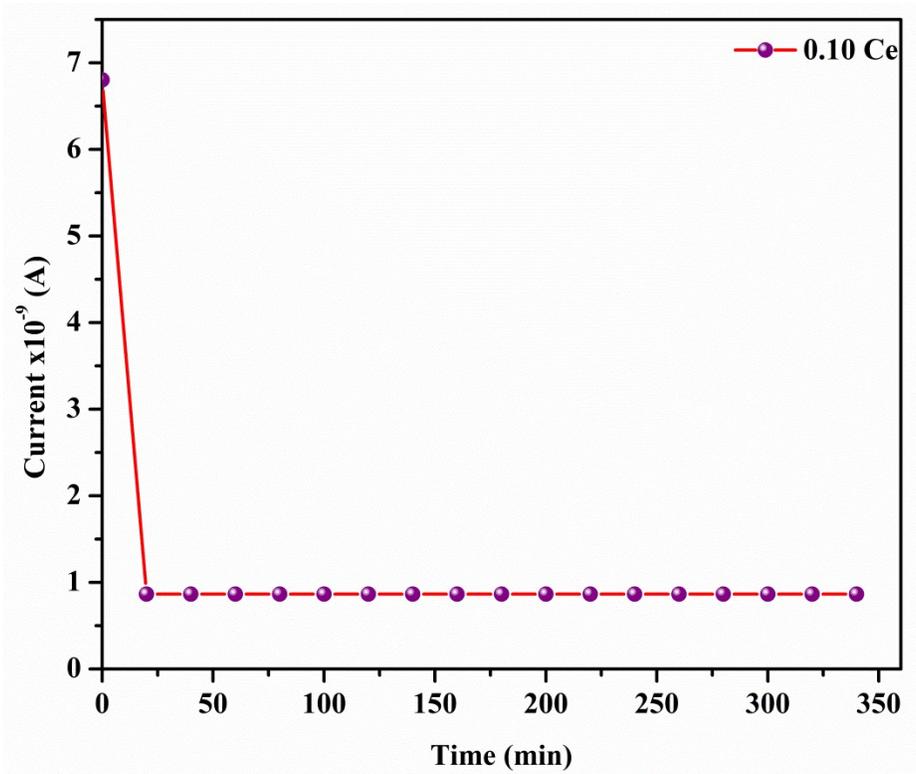

**Fig.9: DC conductivity graph for 0.10 Ce sample.**

## 4. Conclusions

The series $Li_{6.4}Ga_{0.2}La_3Zr_{2-x}Ce_xO_{12}$ with Ce(x) substitution ranging from 0 to 0.30 a.p.f.u. in Ga doped LLZO was synthesized by conventional solid state reaction method. The synthesized samples were studied using various structural and electrochemical characterizations. The necessary condition of formation of cubic phase was confirmed for all the samples by X-ray diffraction analysis. The result of surface micrographs shows the formation of compact and dense microstructure leading to the highest relative density value for 0.10 Ce ceramic sample. This result can be attributed to the synergetic effects of Ga-Ce co-doping in garnet LLZO, where Ga and Ce acted as a stabilizing and sintering agent respectively. Moreover, 0.10 Ce sample shows the minimum intercept on real impedance axis and gave the highest Li-ion conductivity of $4 \times 10^{-4}$ S/cm with minimum activation energy of 0.29 eV. This conductivity results can b due to the insertion of high ionic radii Ce at the Zr site which redistributed the Li ions in the LLZO lattice and helped in the bottleneck tuning of Li-ion migration pathways. Lastly, the predominance of ionic conduction in the prepared samples was studied by DC conductivity measurement and the transport number was found to > 0.999 suggesting the negligible electronic contribution in the total conductivity. Hence, all the obtained results for 0.10 Ce sample makes it a prominent solid electrolyte to be used in ASSLIBs.

**Acknowledgement**

Authors like to acknowledge Department of Physics, VNIT, Nagpur for providing XRD facility governed by DST FIST project number SR/FST/PSI/2017/5(C).

**Funding**

This research did not receive any specific grant from funding agencies in the public, commercial, or not-for-profit sector.

## References

[1]  S. Ramakumar, C. Deviannapoorani, L. Dhivya, L.S. Shankar, R. Murugan, Lithium garnets: Synthesis, structure, Li+ conductivity, Li+ dynamics and applications, Prog.

Mater. Sci. 88 (2017) 325–411. https://doi.org/10.1016/j.pmatsci.2017.04.007.

[2]  S. Guo, Y. Sun, A. Cao, Garnet-type Solid-state Electrolyte Li7La3Zr2O12: Crystal Structure, Element Doping and Interface Strategies for Solid-state Lithium Batteries, Chem. Res. Chinese Univ. 36 (2020) 329–342. https://doi.org/10.1007/s40242-020-0116-0.

[3]  M. Aote, A. V. Deshpande, Enhancement in Li-ion conductivity through Co-doping of Ge and Ta in garnet Li7La3Zr2O12 solid electrolyte, Ceram. Int. 49 (2023) 40011–40018. https://doi.org/10.1016/j.ceramint.2023.09.330.

[4]  M. Aote, A. V. Deshpande, Effect of Ca doping on Li ion conductivity of Ge and Ta doped garnet LLZO, J. Phys. Chem. Solids. 190 (2024) 111980. https://doi.org/10.1016/j.jpcs.2024.111980.

[5]  M. Aote, A. V. Deshpande, Study of Ge-doped garnet type Li7La3Zr2O12 as solid electrolyte for Li-ion battery application, J. Mater. Sci. Mater. Electron. 35 (2024). https://doi.org/10.1007/s10854-024-12338-5.

[6]  V. Thangadurai, S. Adams, W. Weppner, Crystal structure revision and identification of Li+-ion migration pathways in the garnet-like Li5La3M 2O12 (M = Nb, Ta) oxides, Chem. Mater. 16 (2004) 2998–3006. https://doi.org/10.1021/cm031176d.

[7]  R. Murugan, V. Thangadurai, W. Weppner, Fast lithium ion conduction in garnet-type Li7La 3Zr2O12, Angew. Chemie - Int. Ed. 46 (2007) 7778–7781. https://doi.org/10.1002/anie.200701144.

[8]  Y. Cao, Y.Q. Li, X.X. Guo, Densification and lithium ion conductivity of garnet-type Li 7-xLa3Zr2-xTaxO12 (x = 0.25) solid electrolytes, Chinese Phys. B. 22 (2013). https://doi.org/10.1088/1674-1056/22/7/078201.

[9]  J.L. Allen, J. Wolfenstine, E. Rangasamy, J. Sakamoto, Effect of substitution (Ta, Al, Ga) on the conductivity of Li 7La 3Zr 2O 12, J. Power Sources. 206 (2012) 315–319. https://doi.org/10.1016/j.jpowsour.2012.01.131.

[10] A.R. Yoo, S.A. Yoon, Y.S. Kim, J. Sakamoto, H.C. Lee, A comparative study on the

synthesis of Al-doped Li6.2La3Zr2O12 powder as a solid electrolyte using sol-gel synthesis and solid-state processing, J. Nanosci. Nanotechnol. 16 (2016) 11662–11668. https://doi.org/10.1166/jnn.2016.13570.

[11] S. Aktaş, O.M. Özkendir, Y.R. Eker, Ş. Ateş, Ü. Atav, G. Çelik, W. Klysubun, Study of the local structure and electrical properties of gallium substituted LLZO electrolyte materials, J. Alloys Compd. 792 (2019) 279–285. https://doi.org/10.1016/j.jallcom.2019.04.049.

[12] C. Li, A. Ishii, L. Roy, D. Hitchcock, Y. Meng, K. Brinkman, Solid-state reactive sintering of dense and highly conductive Ta-doped Li7La3Z2O12 using CuO as a sintering aid, J. Mater. Sci. 55 (2020) 16470–16481. https://doi.org/10.1007/s10853-020-05221-1.

[13] D.K. Schwanz, A. Villa, M. Balasubramanian, B. Helfrecht, E.E. Marinero, Bi aliovalent substitution in Li7La3Zr2O12 garnets: Structural and ionic conductivity effects, AIP Adv. 10 (2020). https://doi.org/10.1063/1.5141764.

[14] S. Song, D. Sheptyakov, A.M. Korsunsky, H.M. Duong, L. Lu, High Li ion conductivity in a garnet-type solid electrolyte via unusual site occupation of the doping Ca ions, Mater. Des. 93 (2016) 232–237. https://doi.org/10.1016/j.matdes.2015.12.149.

[15] D. Zhao, Y. Yang, C. Zhang, G. Zhu,  Effect of Nb Doping on Phase, Microstructure and Lithium-ion Conductivity of Li 7 La 3 Zr 2 O 12 Solid Electrolyte , Eng. Adv. 3 (2023) 84–88. https://doi.org/10.26855/ea.2023.04.001.

[16] C. Shao, Z. Yu, H. Liu, Z. Zheng, N. Sun, C. Diao, Enhanced ionic conductivity of titanium doped Li7La3Zr2O12 solid electrolyte, Electrochim. Acta. 225 (2017) 345–349. https://doi.org/10.1016/j.electacta.2016.12.140.

[17] K.B. Dermenci, A.F. Buluç, S. Turan, The effect of limonite addition on the performance of Li7La3Zr2O12, Ceram. Int. 45 (2019) 21401–21408. https://doi.org/10.1016/j.ceramint.2019.07.128.

[18] X. Liang, X. Li, X. Wu, Q. Gai, X. Zhang, Effect of Zn-Incorporation into Li7La3Zr2O12

Solid Electrolyte on Structural Stability and Electrical Conductivity: First-Principles Study, Int. J. Electrochem. Sci. 15 (2020) 11123–11136. https://doi.org/10.20964/2020.11.38.

[19] M. Aote, A. V. Deshpande, A. Khapekar, K. Parchake, Influence of Ga-Ge doping on the structural and electrical properties of Li7La3Zr2O12 solid electrolytes, Mater. Lett. 377 (2024) 137462. https://doi.org/10.1016/j.matlet.2024.137462.

[20] S.H. Yang, M.Y. Kim, D.H. Kim, H.Y. Jung, H.M. Ryu, J.H. Han, M.S. Lee, H.S. Kim, Ionic conductivity of Ga-doped LLZO prepared using Couette–Taylor reactor for all-solid lithium batteries, J. Ind. Eng. Chem. 56 (2017) 422–427. https://doi.org/10.1016/j.jiec.2017.07.041.

[21] J. Wang, X. Li, X. Wang, G. Liu, W. Yu, X. Dong, J. Wang, The synergistic effects of dual substitution of Bi and Ce on ionic conductivity of Li7La3Zr2O12 solid electrolyte, Ceram. Int. 50 (2024) 6472–6480. https://doi.org/10.1016/j.ceramint.2023.11.391.

[22] E. Rangasamy, J. Wolfenstine, J. Allen, J. Sakamoto, The effect of 24c-site ( A ) cation substitution on the tetragonal e cubic phase transition in Li 7 À x La 3 À x A x Zr 2 O 12 garnet-based ceramic electrolyte, 230 (2013) 261–266.

[23] M. Nasir, J. Seo, J.S. Park, H.J. Park, Excellent electrochemical response of Ce stabilized cubic Li7La3Zr2O12, J. Eur. Ceram. Soc. 44 (2024) 4606–4611. https://doi.org/https://doi.org/10.1016/j.jeurceramsoc.2024.01.082.

[24] D. Liu, Y. Hou, C. Bulin, R. Zhao, B. Zhang, Enhancing ionic conductivity of garnet-type Nb-doped Li7La3Zr2O12 by cerium doping, J. Rare Earths. 42 (2024) 1740–1746. https://doi.org/10.1016/j.jre.2023.07.012.

[25] B. Dong, S.R. Yeandel, P. Goddard, P.R. Slater, Combined Experimental and Computational Study of Ce-Doped La3Zr2Li7O12 Garnet Solid-State Electrolyte, Chem. Mater. 32 (2020) 215–223. https://doi.org/10.1021/acs.chemmater.9b03526.

[26] C. Zheng, Y. Lu, J. Su, Z. Song, T. Xiu, J. Jin, M.E. Badding, Z. Wen, Grain Boundary Engineering Enabled High-Performance Garnet-Type Electrolyte for Lithium Dendrite

Free Lithium Metal Batteries, Small Methods. 6 (2022) 1–10. https://doi.org/10.1002/smtd.202200667.

[27] X. Huang, C. Liu, Y. Lu, T. Xiu, J. Jin, M.E. Badding, Z. Wen, A Li-Garnet composite ceramic electrolyte and its solid-state Li-S battery, J. Power Sources. 382 (2018) 190–197. https://doi.org/10.1016/j.jpowsour.2017.11.074.

[28] Z. Xu, X. Hu, B. Fu, K. Khan, J. Wu, T. Li, H. Zhou, Z. Fang, M. Wu, Co-doping strategy enhanced the ionic conductivity and excellent lithium stability of garnet-type $Li_7La_3Zr_2O_{12}$ electrolyte in all solid-state lithium batteries, J. Mater. 9 (2023) 651–660. https://doi.org/10.1016/j.jmat.2023.01.007.

[29] N. Birkner, C. Li, S.L. Estes, K.S. Brinkman, Gallium-Doping Effects on Structure, Lithium-Conduction, and Thermochemical Stability of $Li_{7-3x}Ga_xLa_3Zr_2O_{12}$ Garnet-Type Electrolytes, ChemSusChem. 14 (2021) 2621–2630. https://doi.org/10.1002/cssc.202100526.

[30] Y. Chen, E. Rangasamy, C. Liang, K. An, Origin of High $Li^+$ Conduction in Doped $Li_7La_3Zr_2O_{12}$ Garnets, Chem. Mater. 27 (2015) 5491–5494. https://doi.org/10.1021/acs.chemmater.5b02521.

[31] M. Aote, A. V Deshpande, K. Parchake, A. Khapekar, Effect of Sr on the Ionic Conductivity of Ta Doped Garnet $Li_7La_3Zr_2O_{12}$ Solid Electrolyte, ArXiv E-Prints. (2024) arXiv--2411.